\documentclass[sigchi, nonacm]{acmart}
\AtBeginDocument{%
  }

\usepackage{tcolorbox}

\begin{document}

\title{Utilizing ChatGPT in a Data Structures and Algorithms Course: A Teaching Assistant's Perspective}

\author{Pooriya Jamie}
\email{Pooriya.Jamie@gmail.com}
\orcid{0009-0008-5750-111X}
\affiliation{%
  \institution{Amirkabir University of Technology (Tehran Polytechnic)}
  \city{Tehran}
  \country{Iran}
}

\author{Reyhaneh Hajihashemi}
\email{reyhanehhashemi171@gmail.com}
\orcid{0009-0005-7450-1355}
\affiliation{%
  \institution{Amirkabir University of Technology (Tehran Polytechnic)}
  \city{Tehran}
  \country{Iran}
}

\author{Sharareh Alipour}
\email{sharareh.alipour@gmail.com}
\orcid{0000-0002-3626-8960}
\affiliation{%
  \institution{Tehran Institute for Advanced Studies (TEIAS)}
  \city{Tehran}
  \country{Iran}
}

\renewcommand{\shortauthors}{P. Jamie, R. Hajihashemi, and S. Alipour}

\renewcommand{\abstractname}{ABSTRACT}

\begin{abstract}
Integrating large language models (LLMs) like ChatGPT into computer science education offers transformative potential for complex courses such as data structures and algorithms (DSA). This study examines ChatGPT as a supplementary tool for teaching assistants (TAs), guided by structured prompts and human oversight, to enhance instruction and student outcomes. A controlled experiment compared traditional TA-led instruction with a hybrid approach where TAs used ChatGPT-4o and ChatGPT o1 to generate exercises, clarify concepts, and provide feedback. Structured prompts emphasized problem decomposition, real-world context, and code examples, enabling tailored support while mitigating over-reliance on AI. Results demonstrated the hybrid approach’s efficacy, with students in the ChatGPT-assisted group scoring 16.50 points higher on average and excelling in advanced topics. However, ChatGPT’s limitations necessitated TA verification. This framework highlights the dual role of LLMs: augmenting TA efficiency while ensuring accuracy through human oversight, offering a scalable solution for human-AI collaboration in education.
\end{abstract}



\maketitle

% In the document body
\section{INTRODUCTION}
Integrating large language models (LLMs) such as ChatGPT into educational settings is transforming how we approach teaching and learning \cite{wang2024largelanguagemodelseducation, 10.1371/journal.pone.0304013}. LLMs can significantly improve student engagement by providing instant feedback, enhance understanding by breaking down complex concepts, and boost performance by aiding exam preparation \cite{10.1145/3657604.3662036, 10.1145/3636243.3636248}. Despite the growing interest in LLMs, there is still a significant gap in the literature regarding how teaching assistants (TAs) can effectively utilize these tools to enhance their teaching and address student questions, particularly in data structures and algorithms (DSA) courses. While much of the existing research focuses on students' direct interactions with LLMs or using LLMs as standalone TAs, there has been limited exploration of how TAs can leverage these technologies to improve their teaching and provide better support for students.

DSA courses are fundamental to computer science education, playing a crucial role in developing students' understanding of algorithms, programming skills, and problem-solving strategies \cite{Barczak2023}. However, the complexity of these courses often presents significant challenges for students, underscoring the need for innovative teaching aids. While LLMs like ChatGPT offer promising solutions, their unsupervised use can lead to challenges such as academic integrity issues and over-reliance on AI-generated content \cite{doi:10.1080/14703297.2023.2190148}.

In this study, we explore how TAs can use ChatGPT as a supplementary tool to assist in answering student questions during TA classes or office hours. Specifically, we investigate the following research questions:

\begin{itemize} 
\item \textbf{RQ1:} How does the use of ChatGPT by TAs impact students’ learning and exam readiness compared to traditional TA-led instruction?
\\
\item \textbf{RQ2:} What challenges arise in utilizing ChatGPT to answer DSA questions? 
\\
\item \textbf{RQ3:} How can the different versions of ChatGPT (4o and o1) overcome limitations and support each other to enhance student learning outcomes?
\end{itemize}

To answer these questions, we conducted a study in a DSA course, where TAs used ChatGPT as an assistant, guided by structured prompts, to answer student questions. Two versions of ChatGPT, ChatGPT-4o and ChatGPT o1, were used to support different tasks, with TAs verifying the accuracy of the responses. This approach mitigates the limitations of using LLMs independently and provides a scalable solution for AI integration in education.

\section{RELATED WORK AND OUR CONTRIBUTION}
\subsection{LLMs in Computer Science Education}

Despite the growing interest, there is a notable gap in research focusing specifically on how LLMs can be effectively utilized by teaching assistants (TAs) in advanced computer science courses, particularly data structures and algorithms (DSA). While much of the existing literature evaluates LLMs as general learning aids, the specific challenges and opportunities associated with TA-guided LLM use in complex subject areas, such as providing structured guidance and addressing misconceptions, are underexplored.

Research studies examining LLMs' impact on education typically fall into two categories: those that highlight the benefits of LLMs as effective educational tools, and those that point out their drawbacks, considering LLMs as potentially detrimental in educational settings \cite{10.1145/3587102.3588852, Leinonen_2023_using, Leinonen_2023_Comparing, 10.1145/3501385.3543957, 10.1145/3613905.3637148, KASNECI2023102274, 10.1145/3631802.3631806}.

Denny et al. \cite{10.1145/3649217.3653574} highlights LLMs as programming teaching aids, valuing feedback, problem-solving, and autonomy, while addressing access and over-reliance challenges.
Farrokhnia et al. \cite{doi:10.1080/14703297.2023.2195846} provide a comprehensive analysis of the strengths, weaknesses, opportunities, and threats (SWOT) associated with the use of ChatGPT in educational settings.
Yiyin Shen et al. \cite{10.1145/3626252.3630874} found that prompt engineering could improve ChatGPT's effectiveness in data science assignments, particularly when performance was initially low, highlighting the role of LLMs and the importance of prompt engineering in education.

\subsection{Impact on Learning Outcomes}

The integration of LLMs in education has been linked to improvements in student engagement and academic performance \cite{10.1145/3613905.3650868}. For example, Kwan Lo \cite{educsci13040410} demonstrated that ChatGPT can personalize learning and automate administrative tasks, resulting in more effective online learning environments. Singh et al. \cite{educsci13090924} further demonstrated that ChatGPT enhances student satisfaction and academic achievements due to the accessible academic support it provides. Dempere et al. \cite{10.3389/feduc.2023.1206936} also discussed ChatGPT's role in supporting research and automating grading, which improves teaching methodologies and student retention.

However, these benefits are counterbalanced by notable concerns, such as potential academic dishonesty, reduced creativity, diminished human interaction, and the risk of LLMs providing incorrect information \cite{mohan2024analysis, 10.3389/feduc.2023.1206936, educsci13090924, Perkins2023}. Additionally, some articles discuss the possibility of LLMs acting as replacements for teachers \cite{anishka2024chatgpt}, which raises concerns about the appropriateness and limitations of these technologies in classrooms.

\subsection{LLMs in Data Structures and Algorithms}

The impact of LLMs in advanced computer science subjects, such as DSA, has received significant attention. Prior research has primarily focused on the effectiveness of LLMs as standalone tools for supporting student learning and helping with coding tasks \cite{anagnostopoulos2023chatgpt, 10.1145/3626252.3630803}. These studies indicate that LLMs can increase productivity, particularly for beginners, by assisting with coding, programming, and understanding algorithmic concepts. However, concerns remain regarding potential errors in generated code and an incomplete understanding of programming fundamentals.

Denny et al. \cite{10.1145/3545945.3569823} assessed Copilot's accuracy in solving various Python problems, noting that while it solved many problems effectively, it also struggled with more complex and conceptual challenges. Similarly, Dakhel et al. \cite{MORADIDAKHEL2023111734} found that while Copilot generated solutions for sorting algorithms and basic graph problems, the quality of these solutions did not consistently match that produced by human students in terms of accuracy and complexity.

These findings highlight the potential and limitations of LLMs in computer science education, implying the necessity of guided integration with structured prompts and TA support to maximize benefits and mitigate drawbacks.

\subsection{Our Contribution}

This study explores how TAs can use ChatGPT as a supplementary tool to enhance their teaching in a DSA course. The novelty lies in the combination of TA expertise + ChatGPT, guided by structured prompts, to answer student questions during TA classes or office hours. This approach addresses challenges in DSA education, such as over-reliance on AI and comprehension gaps, by ensuring that students receive reliable, TA-verified information while reducing the workload on TAs. By leveraging ChatGPT, TAs can focus on more complex teaching aspects, improving both efficiency and educational quality.
Unlike previous studies that primarily examined AI as a direct tutor or a standalone tool for students, our work emphasizes the synergy between human oversight and AI guidance, ensuring that the TA remains an active, pedagogically informed mediator. In our approach, the TA not only curates and validates AI-generated content but also integrates it into a structured learning process that actively supports and guides students.
Our findings demonstrate how ChatGPT-4o and o1 versions helped create innovative problem sets and clarify complex algorithms, enhancing the educational experience. This approach provides a scalable solution for integrating AI into education while maintaining the quality and reliability of TA-led instruction.

\section{METHODOLOGY}

Our study involved 40 undergraduate students in a data structures and algorithms (DSA) course at a top-tier university, with equal male and female representation. Participants were randomly assigned to two groups: one group received answers from the TA using ChatGPT, while the other group received answers from the TA in the traditional way, without the use of ChatGPT. Both groups had similar prior knowledge and study habits, confirmed by pre-tests.

The TA, experienced from 16 related courses, facilitated activities and gave personalized feedback. ChatGPT-4o and ChatGPT o1, directed by structured prompts, were employed to create practice problems, debug code, and improve teaching methods. The TA checked response to ensure effective engagement and tool utilization.

\subsection{ChatGPT-assisted TA instruction}
\subsubsection{Setup and Supervision}

The primary objective of this study was to examine the impact of ChatGPT as an assistant to TAs in answering student questions on their learning performance. We focused on data structures and algorithms (DSA) exercises, addressing the tendency of students to accept AI-generated answers without fully understanding the concepts. To tackle this issue, we conducted an experiment with two groups of students:

\begin{itemize}
    \item \textbf{Traditional TA Instruction Group}: This group received answers from the TA in the traditional way, without the use of ChatGPT. The TA relied on their own knowledge and expertise to answer questions and design in-class exercises.
    
    \item \textbf{TA + ChatGPT Group}: In this group, the TA used ChatGPT, guided by structured prompts, to answer student questions and design in-class exercises. The TA generated both theoretical and practical questions using ChatGPT but verified the accuracy of the answers and questions before providing them to students during TA classes. This combination of TA expertise + ChatGPT ensured that students received reliable, creative, and well-vetted questions and answers, as opposed to traditional TA classes or students using ChatGPT independently.
\end{itemize}

In the TA + ChatGPT Group, the TA actively mediated the process by critically evaluating and refining ChatGPT’s outputs in real time, ensuring that every answer and question met rigorous pedagogical standards before being presented to students.

The structured prompts were designed to break down problems into manageable parts, guiding students step-by-step toward the solution while maintaining clarity and relevance. These prompts were iteratively refined over three rounds based on feedback from students and the TA. The refinement process focused on improving clarity, relevance, and problem-solving guidance. For example, students indicated that the prompts needed clearer guidance on common pitfalls, which led us to add specific warnings and additional explanations for challenging steps.
The final prompt version included the following elements:

\begin{itemize}
\item \textbf {Problem Understanding:} Clarifying the problem statement.
\item \textbf {Key Characteristics:} Identifying the essential properties of the problem.
\item \textbf {Common Operations:} Discuss typical operations related to the problem.
\item \textbf {Algorithm Description:} Outlining potential algorithms for solving the problem.
\item \textbf {Real-World Scenario:} Providing context to relate the problem to real-life applications.
\item \textbf {Code Snippet:} Offering a basic code structure.
\end{itemize}

Figure \hyperref[fig:prompt]{1} shows the TA-prepared prompt designed to break down problems into manageable parts and guide students step-by-step toward the solution. Key elements like problem understanding, algorithm description, and real-world scenarios were particularly effective in enhancing student comprehension. By using structured prompts, the TA ensured that students engaged critically with the material rather than passively accepting AI-generated answers.

\begin{figure}[htbp]
    \begin{tcolorbox}[colback=gray!30,colframe=black,rounded corners]
    \textbf{Prompt:} \textit{"Provide a comprehensive step-by-step explanation of the given data structures and algorithms problem, focusing on its key characteristics, common operations, and algorithm description. Break down the problem into smaller, manageable steps and guide the reader through each step, encouraging them to think critically and reason through the solution. Along the way, explain key concepts, relevant techniques, and any necessary assumptions. Additionally, illustrate a real-world scenario where these data structures and algorithms can be applied and provide a code snippet showcasing their implementation in a programming language of your choice."}
    \end{tcolorbox}
    \caption{TA-prepared prompt for DSA questions}
    \label{fig:prompt}
    \Description[Prompt for data structures and algorithms question]{This figure presents a text box containing a comprehensive prompt prepared by a teaching assistant for data structures and algorithms (DSA) questions. The prompt reads: "Please provide a comprehensive step-by-step explanation of the given data structures and algorithms problem, focusing on its key characteristics, common operations, and algorithm description. Break down the problem into smaller, manageable steps and guide the reader through each step, encouraging them to think critically and reason through the solution. Along the way, explain key concepts, relevant techniques, and any necessary assumptions. Additionally, illustrate a real-world scenario where these data structures and algorithms can be applied and provide a code snippet showcasing their implementation in a programming language of your choice."}
\end{figure}

\subsubsection{Evaluation}
To evaluate the effectiveness of ChatGPT integration, we established specific, measurable objectives for student satisfaction and exam performance. We used a structured survey at the end of the course to gauge student satisfaction, including how helpful students found ChatGPT, their understanding of complex concepts, and overall satisfaction.

To measure learning outcomes, we compared exam scores of the two groups. We also analyzed their performance on various types of questions to identify improvement areas, such as problem-solving and theoretical understanding.

By incorporating structured prompts and TA oversight, we aimed to create an environment where ChatGPT was used as a supplementary tool for the TA, rather than a shortcut for students. Our comparative study demonstrated that this approach led to better learning outcomes and higher exam scores compared to traditional TA-only instruction.

\subsection{Identifying ChatGPT's Limitations}

Through our experiment, we observed several limitations in ChatGPT's ability to interact with DSA-related questions, such as difficulty generating creative problems and struggles with visual representation and complex algorithmic reasoning. To better understand these challenges, we analyzed ChatGPT's performance on a diverse set of theoretical and practical questions, ranging from basic data structures (arrays, linked lists) to advanced topics (graph traversal algorithms, dynamic programming). The TA used structured prompts to guide ChatGPT in generating responses, which were then verified for accuracy before being shared with students. This process highlighted the importance of human oversight in addressing ChatGPT's limitations.

We assessed ChatGPT's responses across several dimensions. Accuracy was evaluated by analyzing whether ChatGPT provided correct answers, especially for complex DSA problems involving recursion and dynamic programming. Depth of Explanation focused on whether ChatGPT offered thorough, step-by-step explanations for algorithmic problems, ensuring sufficient clarity. When examining Handling Novel Scenarios, we observed ChatGPT's ability to respond to creative questions outside its pre-trained dataset, noting its tendency to provide incomplete or superficial answers for novel problems. Visual Representation and Code Generation were tested to assess ChatGPT’s capability to visually represent problems, such as tree structures or graph traversals, and its ability to generate accurate and logically sound code.
To further investigate ChatGPT's limitations, we collected feedback from students and professors who utilized ChatGPT in their studies and teaching. Their experiences highlighted key challenges, such as the model’s inability to generate creative exercises and its occasional production of code snippets that were syntactically correct but logically flawed, requiring further refinement. These insights emphasized the need for expert oversight when using ChatGPT for complex educational tasks.

\subsection{Enhanced TA Classes Using ChatGPT-4o and o1 Series}

To address the challenges observed in TA-assisted learning with ChatGPT-4o, such as limited reasoning depth, incomplete solutions for complex problems, and insufficient support for advanced algorithmic concepts, we integrated the ChatGPT o1 series, which offers improved problem-solving capabilities. This allowed us to explore how different versions of ChatGPT could complement each other in supporting TA-led learning.

ChatGPT-4o improved speed, multi-modal inputs, and multi-language support, boosting class material creation and real-time feedback. ChatGPT o1 excelled in complex reasoning, aiding recursive algorithms and data structure tasks. This enabled us to explore varied TA-led learning support across versions.

In TA-led sessions, both models were integrated into the classroom activities to maximize their complementary strengths. During the preparation phase, the TA used ChatGPT-4o to generate a diverse set of basic questions from the teacher’s booklet. For more complex problems, ChatGPT o1 was employed to provide detailed explanations and step-by-step reasoning, supplementing ChatGPT-4o’s output.

In interactive sessions, the TA used questions from both models. ChatGPT o1’s chain-of-thought reasoning helped guide students through complex concepts like dynamic programming, promoting engagement and critical thinking.

For troubleshooting and clarification, ChatGPT o1 excelled in breaking down advanced problems into logical steps, while ChatGPT-4o handled image-related tasks under the supervision of the TA to ensure precision and correctness.

During assessments, ChatGPT-4o reviewed student responses for general feedback, while ChatGPT o1 refined this feedback by addressing deeper logical issues. This combination improved TA support and classroom experiences, with ChatGPT-4o handling foundational content and ChatGPT o1 focusing on advanced reasoning.

\section{RESULTS}
\subsection{Student Performance Analysis}

We analyzed the outcomes of ChatGPT-assisted TA instruction, focusing on key metrics such as student performance, engagement, and exam results. Students were divided into two groups.

The Traditional TA Instruction Group received answers directly from the TA without ChatGPT, relying solely on their expertise. In contrast, the Hybrid TA + ChatGPT Group utilized ChatGPT as an assistant with structured prompts, where the TA verified the accuracy of responses and exercises before presenting them to students.
This comparison allowed us to assess the impact of ChatGPT as a supplementary tool for TAs versus traditional TA-only instruction.

\begin{figure*}[htbp]
\centering
\includegraphics[width=\linewidth]{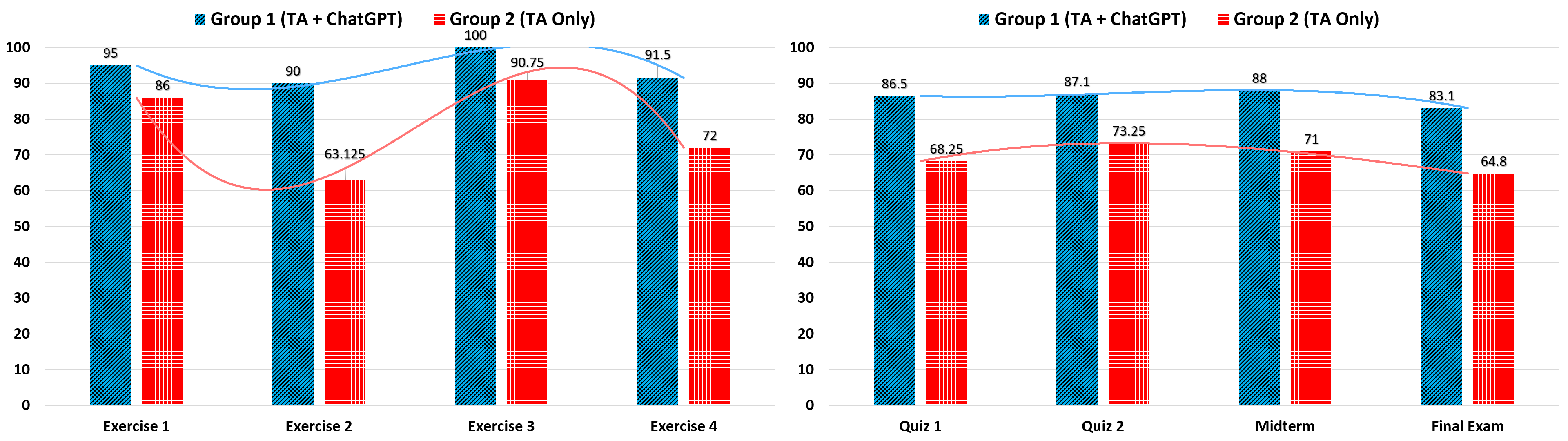}
\caption{Average scores for exercises, quizzes, midterms, and final exams}
\label{fig:merged}
\Description[Average scores for exercises, quizzes, midterms, and final exams]{This figure presents two graphs comparing the average scores between two groups. Group 1 (TA + ChatGPT), represented by a blue line and bars, includes students who used ChatGPT under TA supervision. Group 2 (TA Only), represented by a red line and bars, received traditional TA support without ChatGPT. The TA + ChatGPT group consistently achieved higher scores across exercises, quizzes, midterms, and final exams compared to the TA Only group.}
\end{figure*}

The results are illustrated in Figure \hyperref[fig:merged]{2}, which shows the average scores for exercises, quizzes, midterms, and final exams for both groups. Group 1 (TA + ChatGPT) consistently outperformed Group 2 (TA Only) across all metrics, with a mean score difference of 16.50 ($p < 0.001$). This significant improvement indicates that when TAs use ChatGPT as an assistant, guided by structured prompts, they are able to answer student questions in a more creative, complete, and accurate manner. The structured prompts ensured that no critical information was lost, and the TA's verification process guaranteed the reliability of the responses. As a result, students in the TA + ChatGPT group demonstrated a deeper understanding of complex concepts, particularly in challenging areas such as recursion and dynamic programming, leading to significantly better performance in problem-solving tasks.

The group where TAs used ChatGPT as an assistant consistently outperformed the TA-only group in assignments and exams. Students in the TA-only group showed fluctuating scores, especially on complex exercises, which often resulted in weaker exam performance. In contrast, the ChatGPT-assisted group maintained stable, higher scores, benefiting from the structured prompts and the TA's verification of responses. They excelled in advanced topics, scoring on average 15-20\% higher than the TA-only group. This highlights a key advantage of combining TA expertise with ChatGPT: not only are the responses verified by an expert (the TA in this study), but ChatGPT also provides a creative and robust instructional approach that can complement or even enhance the TA's knowledge. In real-time classroom settings, where even experts may occasionally overlook details, ChatGPT serves as a valuable tool to fill gaps and provide additional insights. The structured prompts and TA oversight clarified problems, provided step-by-step guidance, and offered targeted hints, significantly enhancing students' problem-solving skills and conceptual understanding.

\begin{figure*}[htbp]
\centering
\includegraphics[width=0.8\linewidth]{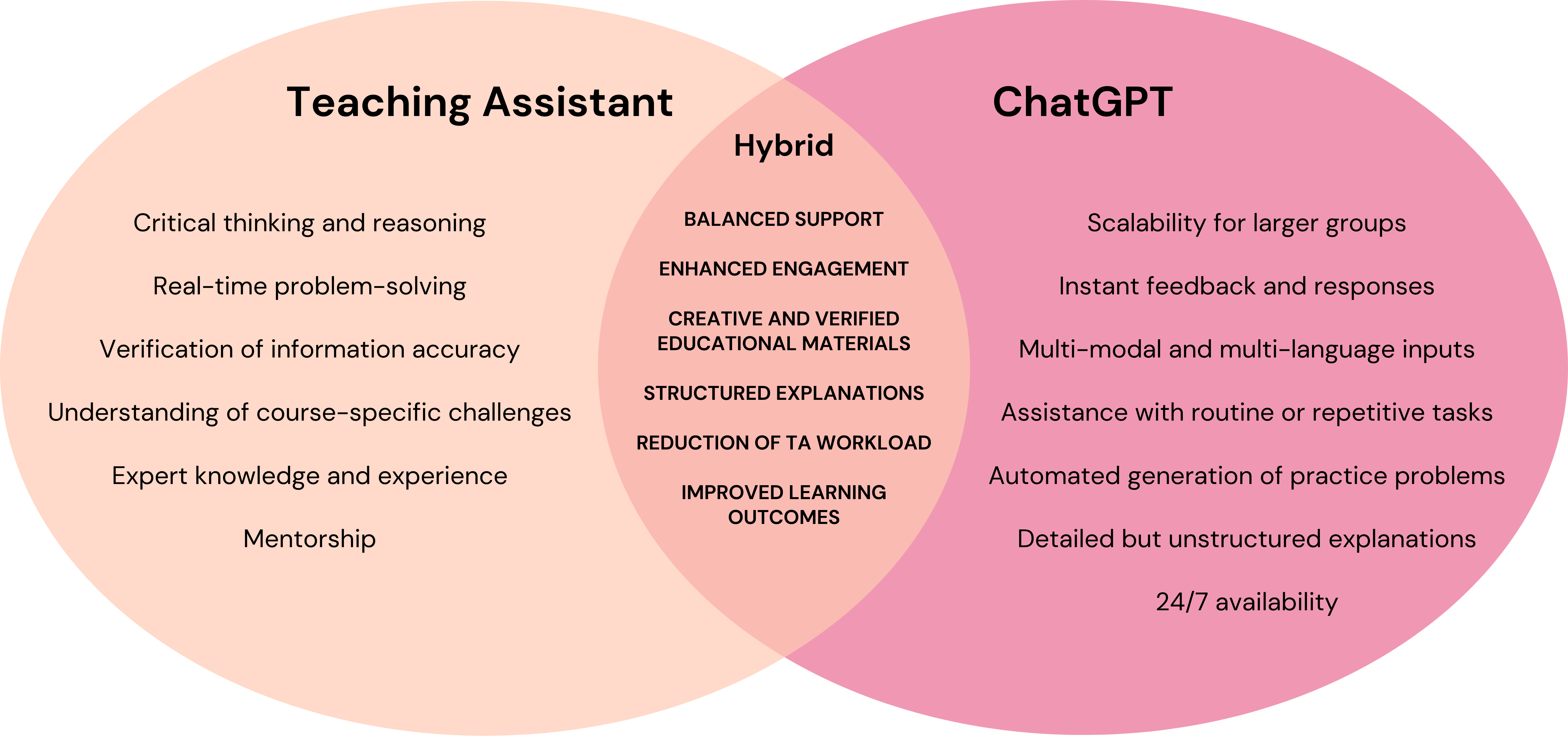}
\caption{Comparison of Features: Teaching Assistants, ChatGPT, and Hybrid Contributions}
\label{fig:venn}
\Description[A Venn diagram comparing the features of TAs and ChatGPT with overlapping contributions]{This Venn diagram highlights the unique features of Teaching Assistants (TAs) and ChatGPT, such as expert knowledge and instant feedback, respectively. The overlap represents the hybrid contributions combining TA expertise and ChatGPT capabilities, leading to enhanced learning outcomes and a structured approach.}
\end{figure*}

To further illustrate the unique contributions of our hybrid approach, we present a Venn diagram (Figure \hyperref[fig:venn]{3}) that highlights the distinct features of TAs and ChatGPT, as well as the intersection where their combined strengths enhance DSA education.

Feedback from students highlighted the usefulness of structured prompts and the importance of active engagement. One student appreciated ChatGPT's explanations when using structured prompts but emphasized the need for TA input on complex questions. Another student found ChatGPT helpful for learning DSA, noting its role as an educational assistant, though not a substitute for a teacher. A third student shared her experience in TA classes, highlighting how a TA's use of ChatGPT helped her reach the answers to the questions step by step in a more thorough and creative way.

These findings support our hypothesis that when TAs use ChatGPT as an assistant, guided by structured prompts, it leads to improved learning outcomes and higher exam performance compared to traditional instruction methods.

\subsection{ChatGPT Limitations}

Our evaluation revealed several limitations in ChatGPT's ability to handle DSA-related questions, especially those requiring creativity, visual representation, and advanced reasoning. ChatGPT struggled with topics like dynamic programming and recursion, often producing incomplete or incorrect solutions. For example, it generated misleading graphs for sorting algorithms and failed to adapt code for specific complexities.

While effective for simpler tasks, ChatGPT's responses to complex problems often lacked depth or clarity. Its explanations for algorithms like Dijkstra's were too abstract, making it difficult for students to apply the concepts. Visual representations, such as balanced binary trees or graph traversals, were frequently inaccurate or incomplete.

Generated code was syntactically correct but prone to logical errors, particularly in advanced tasks. Issues in hashing techniques and optimization problems like the knapsack problem highlighted gaps in its ability to break down complex steps comprehensibly.

Feedback indicated that ChatGPT was helpful for basic queries but struggled with edge cases and optimization in advanced problems, such as linked lists. These findings suggest that while promising for simpler content, ChatGPT requires improvements in creativity, explanation depth, and problem-solving. Human oversight remains essential to ensure the quality and accuracy of its outputs in educational settings.

\subsection{ChatGPT-4o and o1 Integration's Results}

The integration of ChatGPT-4o and o1 significantly enhanced the teaching and learning experience in TA-led DSA courses. ChatGPT-4o effectively supported routine tasks, such as generating diverse class materials and providing real-time feedback. However, it exhibited limitations in handling complex problems, such as incomplete solutions for dynamic programming and graph traversal.
ChatGPT o1, on the other hand, excelled in areas where ChatGPT-4o struggled. Its advanced reasoning capabilities allowed it to provide detailed explanations for complex problems, such as recursive algorithms and dynamic programming. For example, during interactive sessions, ChatGPT o1’s chain-of-thought reasoning guided students through multi-step problems, fostering engagement and critical thinking.

In assessments, ChatGPT-4o provided initial feedback, while ChatGPT o1 refined it by identifying deeper logical issues, such as missing break statements in nested loops. This combination of models created a more dynamic learning environment, allowing for both breadth and depth in the educational process.

\section{DISCUSSION AND FUTURE WORK}

\subsection{Summary and Key Findings}

This study evaluated the impact of integrating ChatGPT as a TA assistant in a data structures and algorithms (DSA) course. The results showed that when TAs used ChatGPT with structured prompts and verified its responses, student engagement and academic performance significantly improved compared to traditional TA-only instruction. This highlights ChatGPT's potential as a valuable support tool in complex courses like DSA, where TAs benefit from additional assistance in addressing challenging questions. The combined approach of TA expertise and ChatGPT ensured students received reliable, creative, and well-verified information, enhancing their understanding of advanced topics such as recursion and dynamic programming.

\subsection{Interpretations, Implications, and Balancing LLMs with Human Instruction}

The hybrid approach of ChatGPT, structured prompts, and TA support fostered critical thinking rather than passive acceptance of AI-generated answers. TAs verified ChatGPT’s responses while structured prompts guided students through problem-solving. This aligns with existing literature but extends it by demonstrating the effectiveness of combining AI tools with human expertise. Unlike previous studies that proposed LLMs as replacements for educators \cite{anishka2024chatgpt, Agarwal2023AIIE}, this research showed the advantages of a balanced model, where ChatGPT supports routine tasks and TAs handle critical thinking, individual needs, and complex instruction. The findings suggest that LLMs can be valuable tools when used alongside human instruction, improving educational outcomes.

This study contributes to understanding LLMs in education by showing how ChatGPT can complement traditional teaching. By automating routine tasks and providing tailored feedback, ChatGPT reduces TA workload, enabling them to focus on more complex instructional activities. This integration enhances student learning and provides a scalable solution for larger classes.

\subsection{Ethical Considerations}

We obtained informed consent from students and debriefed them post-study. Participants were randomly assigned to groups with equal resources, ensuring adherence to ethical guidelines. The study evaluated ChatGPT's role as a TA assistant, with both groups receiving equal TA support, though the impact of ChatGPT was only assessed during TA-led sessions. While independent use of chatbots outside of class could influence results, findings suggest that ChatGPT is an effective TA assistant in a controlled classroom environment.

\subsection{Challenges, Limitations, and Future Directions}

Despite its benefits, ChatGPT has limitations, such as difficulty with visual tasks (e.g., graph/tree structures) and inconsistent responses for complex problems. While ChatGPT o1 improved some aspects, such as reasoning for multi-step problems, challenges like generating novel problem sets and handling creative questions persisted. These limitations highlight the need for ongoing human oversight and refinement of LLMs for educational use.
While our study controlled for factors such as prior knowledge and study habits, it is important to acknowledge that additional variables—such as the TA's teaching style, student motivation, and familiarity with AI tools—may also influence the observed improvements. Future research could systematically examine how these factors interact with the hybrid TA–ChatGPT approach.
Future research should focus on optimizing LLMs integration through improved prompts and better handling of complex problems. Expanding LLMs usage across different subjects and educational levels could improve learning outcomes in DSA and beyond. Additionally, scaling the hybrid model to accommodate larger classes and diverse educational contexts will enhance its applicability. The findings from this DSA course are relevant to other computer science disciplines, where problem-solving and critical thinking are key, such as programming, machine learning, or software engineering.

\begin{acks}

This paper has been accepted for publication in CHI EA '25. The final version is available at \href{https://doi.org/10.1145/3706599.3720291}{https://doi.org/10.1145/3706599.3720291}.

\end{acks}

\balance

\bibliographystyle{plain}

\end{document}